\documentstyle{article}

\begin{document}

\baselineskip=23pt

\begin{flushleft}
{\bf {\Huge 
Distinguishing RBL-like objects and XBL-like objects with the peak emission 
frequency of the overall energy spectrum
}}\\
\vspace{2mm}
{\bf Yi-Ping Qin$^{1,2,3}$, G. Z. Xie$^{1,3}$, and Xue-Tang Zheng$^4$\\

$^{1}$ Yunnan Observatory, Chinese Academy of Sciences, Kunming, Yunnan 650011,
P. R. China

$^{2}$ Chinese Academy of Science-Peking University joint Beijing Astrophysical
Center, Department of Geophysics, Peking University, Beijing 100871, P. R.
China

$^{3}$ Yunnan Astrophysics Center, Department of Physics, Yunnan University,
Kunming, Yunnan 650091, P. R. China

$^{4}$ Department of Physics, Nanjing University of Science and Technology,
Nanjing, Jiangsu 210014, P. R. China
}

\vspace{6mm}
{\bf {\Large Abstract}}\\
\end{flushleft}
We investigate quantitatively how the peak emission frequency of the overall
energy spectrum is at work in distinguishing RBL-like and XBL-like objects.
We employ the sample of Giommi et al. (1995) to study the distribution of BL
Lacertae objects with various locations of the cutoff of the overall energy
spectrum. We find that the sources with the cutoff located at lower
frequency are indeed sited in the RBL region of the $\alpha _{ro}-\alpha
_{ox}$ plane, while those with the cutoff located at higher frequency are
distributed in the XBL region. For a more quantitative study, we employ the
BL Lacertae samples presented by Sambruna et al. (1996), where, the peak
emission frequency, $\nu _p$, of each source is estimated by fitting the
data with a parabolic function. In the plot of $\alpha _{rx}-\log \nu _p$ we
find that, in the four different regions divided by the $\alpha _{rx}=0.75$
line and the $\log \nu _p=14.7$ line, all the RBL-like objects are inside
the upper left region, while most XBL-like objects are within the lower
right region. A few sources are located in the lower left region. No sources
are in the upper right region. This result is rather quantitative. It
provides an evidence supporting what Giommi et al. (1995) suggested:
RBL-like and XBL-like objects can be distinguished by the difference of the
peak emission frequency of the overall energy spectrum. \vspace{6mm}

\section{Introduction}

BL Lacertae objects have been mostly observed and identified in either radio
or X-ray surveys. Radio selected BL Lacertae objects (RBLs) differ from
X-ray selected BL Lacertae objects (XBLs) in many aspects. For example, in
radio band, XBLs tend to have weaker cores and they are less core-dominated
than RBLs (see Perlman \& Stocke 1993), while in optical band, XBLs are less
variable and less polarized than most RBLs (see Jannuzi et al. 1994). Most
importantly, more and more studies confirm the fact that RBLs and XBLs
occupy different regions on the $\alpha _{ro}-\alpha _{ox}$ plane,
suggesting that these two groups of objects have significantly different
overall spectral energy distributions (see, e.g., Ledden \& O'Dell 1985;
Stocke et al. 1985; Giommi et al. 1990). According to Padovani \& Giommi
(1995), no explanation of this fact has been given within the different
viewing angle hypothesis so far. In studying the shape of the radio to X-ray
spectrum of a large sample of BL Lacertae objects, Giommi et al. (1995)
found that RBLs are characterized by an energy spectrum with a sharp cutoff
in the IR-to-optical band while in most of XBLs the turnover is located near
the UV-to-X-ray band or at higher frequencies. By assuming that the
wavelength of the peak of the synchrotron emission changes smoothly and
continuously from X-ray band to infrared band for BL Lacertae objects, the
radio-to-optical-to-X-ray colors of RBLs and XBLs can be reproduced and then
explained easily (Padovani \& Giommi 1995; Urry \& Padovani 1995). Padovani
\& Giommi (1996) discovered from the X-ray spectra of $85$ BL Lacertae
objects that there are significant differences between the two classes of
the objects. For XBL-like objects, the X-ray spectral slope is
anti-correlated with the peak emission frequency of the spectrum, while for
RBL-like objects, there is a weak positive correlation between the two
quantities. They found that the observed differences come mainly from the
location of the peak emission frequency of the energy spectrum. Lamer et al.
(1996) got a similar result in their study of soft X-ray spectra of $74$ BL
Lacertae objects. These imply that the location of the peak emission
frequency of the overall energy spectrum of BL Lacertae objects may be an
important quantity comparable to the radio-to-X-ray spectral index $\alpha
_{rx}$ to distinguish RBL-like and XBL-like objects.

Empirically, the frequency where the cutoff of the overall energy spectrum
of a source occurs is also the peak emission frequency of the spectrum when
all the data from radio to X-ray band are presented and plotted. In this
paper, the term ``the cutoff of the overall energy spectrum'' is used when
we deal with only a few observational data of a source, such as that in
Giommi et al. (1995). When the term ``the peak emission frequency of the
spectrum'' is used, we deal with the entire energy spectrum (from radio to
X-ray) of a source.

\section{The role of the cutoff of the overall spectrum}

In this section, we investigate how the location of the cutoff of the
overall energy spectrum is at work in distinguishing the two classes of
BL Lacertae objects, that of RBL-like and XBL-like. According to the study
of Giommi et al. (1995), we expect that, in the $\alpha _{ro}-\alpha _{ox}$
plane, the BL Lacertae objects with a cutoff in IR-to-optical band and those
with a cutoff in UV-to-X-ray band will be distributed in two distinctive
areas which might be defined as RBL and XBL regions respectively.

There are $121$ sources in the sample of Giommi et al. (1995). In this
sample, there are only $103$ sources with all the values of radio, optical
and X-ray fluxes being available. We find from Table 5 of Giommi et al.
(1995) that, for these $103$ sources, the number of the ones with the cutoff
of the overall energy spectrum located at $2.4\times 10^{17}Hz\leq \nu
_p\leq 4.8\times 10^{17}Hz$ is $36$, while that of $5.5\times 10^{14}Hz\leq
\nu _p\leq 6.8\times 10^{14}Hz$ is $60$ and that of $5.0\times 10^{12}Hz\leq
\nu _p\leq 2.5\times 10^{13}Hz$ is $7$. We assign $f_{5GHz}$, $f_{5500\AA }$
and $f_{2keV}$ to radio, optical and X-ray fluxes, respectively. When
available, mean values of these fluxes are adopted. When converting other
X-ray fluxes to $f_{2keV}$, we follow Giommi et al. (1995) to take $\alpha
_x=1.0$ ($f_\nu \propto \nu ^{-\alpha }$). The spectral indices $\alpha _{ro}
$, $\alpha _{ox}$ and $\alpha _{rx}$ for these sources are calculated with $%
\alpha _{ro}=\log (f_{5GHz}/f_{5500\AA })/5.04$, $\alpha _{ox}=\log
(f_{5500\AA }/f_{2keV})/2.94$ and $\alpha _{rx}=\log (f_{5GHz}/f_{2keV})/7.99
$, respectively. 

The K-correction is ignored due to the fact that the effect is small. There
are $71$ of the $103$ sources with their redshifts being available and
certain. We adopt $\alpha _r=0.35$, $\alpha _o=0.5$ and $\alpha _x=1.0$ to
calculate the K-corrected spectral indices $\alpha _{ro}^k$, $\alpha _{ox}^k$
and $\alpha _{rx}^k$ for these $71$ sources. Defining $\triangle \alpha
_{ij}\equiv \alpha _{ij}-\alpha _{ij}^k$, we find $(\triangle \alpha
_{ro})_{\max }=0.009$, $(\triangle \alpha _{ox})_{\max }=0.049$ and $%
(\triangle \alpha _{rx})_{\max }=0.023$ among these sources. It indicates
that, for the above sample, the effect would not significantly affect the
location of a source in the $\alpha _{ro}-\alpha _{ox}$ plane (see Figure 1).

Figure 1 shows that, in the $\alpha _{ro}-\alpha _{ox}$ plane, the $7$
sources with $5.0\times 10^{12}Hz\leq \nu _p\leq 2.5\times 10^{13}Hz$ are
located in the RBL region (where $\alpha _{rx} > 0.75$)
and the $36$ sources with $2.4\times 10^{17}Hz\leq
\nu _p\leq 4.8\times 10^{17}Hz$ are distributed in the XBL region
(where $\alpha _{rx} \leq 0.75$), while the 
$60$ sources with $5.5\times 10^{14}Hz\leq \nu _p\leq 6.8\times 10^{14}Hz $
are scattered in both RBL and XBL regions. Meeting exactly what we expect,
the sources with the cutoff located at lower frequency are indeed sited in
the RBL region while those with the cutoff located at higher frequency are
distributed in the XBL region. Recalling that there are no sufficient data
within optical-to-X-ray band presented in the sample, the scatter of the $60$
sources with $5.5\times 10^{14}Hz\leq \nu _p\leq 6.8\times 10^{14}Hz$ is
understandable. Many of them may probably be those with the cutoff located
in UV band. It is probable that, when sufficient data in UV-to-soft-X-ray
band are available, the above scatter will decrease.

Padovani \& Giommi (1996) pointed out that, for XBL-like objects, $\alpha _x$
is correlated with $\alpha _{ox}$ with a best fit $\alpha _x=(1.38\pm
0.20)\alpha _{ox}-(0.01\pm 0.22)$, while for RBL-like objects, the two
quantities are not correlated. As mentioned above, when converting other
X-ray fluxes to $f_{2keV}$ to calculate $\alpha _{ox}$ we adopt $\alpha
_x=1.0$. Would this significantly affect the distribution of the objects
shown in Figure 1? Among the XBL-like objects (defined by 
$\alpha _{rx} \leq 0.75$) 
in the figure, the largest value of $\alpha _{ox}$ is $1.34$ and the
smallest one is $0.488$. According to the relation of $\alpha _x$ and $%
\alpha _{ox}$ got by Padovani \& Giommi (1996), the two values of $\alpha
_{ox}$ correspond to $\alpha _{x,max}=1.85$ and $\alpha _{x,min}=0.673$, 
respectively.
When converting $f_{1keV}$ to $f_{2keV}$ by adopting $\alpha _x=1.0$, the
corresponding errors $\triangle \alpha _{ox}\equiv \alpha _{ox}-\alpha
_{ox,0}$ (where $\alpha _{ox}$ is valued by adopting the real value of $%
\alpha _x$ and $\alpha _{ox,0}$ is valued by adopting $\alpha _x=1.0$) would
be $-0.087$ and $0.033$, respectively. It suggests that, in Figure 1, when
corrected by this effect, the rightward XBL-like objects would move slightly
leftward and the leftward ones would move slightly rightward. It would not
significantly affect the distribution of the objects.

\section{The plot of the radio-to-X-ray index versus the peak emission
frequency}

Sambruna et al. (1996) collected nonsimultaneous radio, infrared, optical
and X-ray fluxes for three complete samples of blazars to study the
properties of their spectra. In their study, the peak emission frequency of
each source is estimated by fitting the data of the source with a parabolic
function. The values so obtained must more or less reflect the real values
of the peak emission frequencies, since the estimation comes from fitting.
An advantage of these values is that they may be slightly different from
source to source, unlike those of the cutoff of the overall energy spectrum
obtained directly from Table 5 of Giommi et al. (1995), where, only a few
separate values are available. For this reason, these values may be useful
for a more quantitative study. The values of the radio-to-X-ray index $%
\alpha _{rx}$ and the peak emission frequency $\nu _p$ for the three samples
are presented in Table 2 of Sambruna et al. (1996). All these values have
been K-corrected. We adopt only the RBL and XBL samples to study in this
paper. The plot of $\alpha _{rx}-\log \nu _p$ is shown in Figure 2. In
Figure 4 of Sambruna et al. (1996), the solid line is the parabolic fit to
the individual spectral energy distributions including the X-ray data point
and the dashed line is the parabolic fit excluding this point. Among the
sources with dashed lines, two are those with their parabolic lines being
above the X-ray data points and the others are those with their parabolic
lines being under the X-ray data points. We use the filled, empty and empty
plus cross symbols to represent the sources with solid lines, the dashed
lines under the X-ray data points and the dashed lines above the X-ray data
points, respectively, in Figure 2. In this figure, the $\alpha _{rx}=0.75$
line and the $\log \nu _p=14.7$ line divide the $\alpha _{rx}-\log \nu _p$
plane into four different regions. All the RBL-like objects, defined by $%
\alpha _{rx}>0.75$, are inside the upper left region, while most XBL-like
objects, defined by $\alpha _{rx}\leq 0.75$, are within the lower right
region. Those in the lower left region are all the sources with their
parabolic lines being under the X-ray data points in Figure 4 of Sambruna et
al. (1996). There are no sources in the upper right region. Excluding all
the sources with their fitted parabolic lines not passing through the X-ray
data points, those are represented by empty and empty plus cross symbols, we
find from Figure 2 that RBL-like objects and XBL-like objects are well
separated by the $\alpha _{rx}=0.75$ line and the $\log \nu _p=14.7$ line,
and well confined in two different regions. It is known that three points of
data correspond to a single parabolic line. In the plots drawn in Figure 4
of Sambruna et al. (1996), given two leftward points of data, $(\nu _r,\nu
_rL_r)$ and $(\nu _o,\nu _oL_o)$, a larger value of $\nu _xL_x$ for the
rightward point of data, $(\nu _x,\nu _xL_x)$, would yield a larger value
of $\nu _p$ than a smaller value of $\nu _xL_x$ does. Therefore, when
corrected by adjusting the parabolic lines passing through the X-ray data
points, all the empty symbols in Figure 2 will move rightward and all the
empty plus cross symbols will certainly move leftward (a detailed study of
this issue is in preparation). The above result is rather quantitative. It
provides an evidence supporting what Giommi et al. (1995) suggested:
RBL-like and XBL-like objects can be distinguished by the difference of the
peak emission frequency of the overall energy spectrum.

\section{Discussion and conclusions}

To investigate quantitatively how the peak emission frequency of the overall
energy spectrum is at work, we employ the estimated values of the quantity
to study. It shows that, in the plot of $\alpha _{rx}-\log \nu _p$, RBL-like
objects and XBL-like objects are well separated by the two lines of $\alpha
_{rx}=0.75$ and $\log \nu _p=14.7$. The two kinds of objects are well
confined in two different regions. Is this result trivial? Can the value of $%
\nu _p$ be solely determined by $\alpha _{rx}$? Our answer is no. Given only
the values of radio and X-ray fluxes, $f_r$ and $f_x$, $\alpha _{rx}$ is
uniquely determined, but $\nu _p$ is not. It is because that there are
countless parabolic lines passing through $f_r$ and $f_x$. Therefore, $%
\alpha _{rx}$ can not solely determine the value of $\nu _p$. To yield a
single parabolic line, one needs all the values of $f_r$, $f_o$ and $f_x$,
where $f_o$ is the optical flux, [see Equation (2) of Sambruna et al. 1996].
We then reach the conclusion that the above result is not trivial. In fact,
given a certain value of $\alpha _{rx}$, different $\nu _p $ will correspond
to different $f_o$. The quantity $\nu _p$ reflects the relation between $f_r$
and $f_o$ or $f_o$ and $f_x$.

We notice that the value of $\nu _p$ determined by a parabolic function
fitting the values of $f_r$, $f_o$ and $f_x$ is not necessarily the real
value of the peak frequency of the synchrotron emission. This is because
that the shape of the overall energy spectrum is not simply a parabolic line
passing through $f_r$, $f_o$ and $f_x$ for many sources. Their overall
spectra may be contributed by both synchrotron emission and Compton
scattering of lower-energy seed photons (see Figure 6 of Ulrich et al.
1997). Therefore, we do not consider the peak emission frequency $\nu _p$ in
Sambruna et al. (1996) as the real peak frequency of the synchrotron
emission. Instead, we think of the frequency $\nu _p$ as a measure of the
relation between $f_r$ and $f_o$ or $f_o$ and $f_x$. For such a measurement,
we prefer $\nu _p$ to $\alpha _{ro}$ and $\alpha _{ox}$ due to the fact
that, for RBL-like and XBL-like objects, the domains of $\nu _p$ are well
separated, while the domains of $\alpha _{ro}$, as well as $\alpha _{ox}$,
are overlapped (see, e.g., Figure 12 of Padovani \& Giommi 1995).

We then come to the following conclusions. (1) This paper gives a rather
quantitative result which provides an evidence supporting what Giommi et al.
(1995) suggested: RBL-like and XBL-like objects can be distinguished by the
difference of the peak emission frequency of the overall energy spectrum.
(2) The peak emission frequency of a fitted parabolic function may be an
essential quantity comparable to the radio-to-X-ray spectral index $\alpha
_{rx}$ in distinguishing RBL-like and XBL-like objects. The division line is
probably the $\log \nu _p=14.7$ line. (3) The $\alpha _{rx}=0.75$ line
together with the $\log \nu _p=14.7$ line are found to divide the $\alpha
_{rx}-\log \nu _p$ plane into four different regions, where RBL-like objects
are well inside the upper left region and most XBL-like objects are within
the lower right region.

So, we agree with the suggestion of Padovani \& Giommi (1995) that the names
for the two BL Lac classes, radio-selected and X-ray-selected BL Lacs
(RBLs and XBLs), be
replaced by the more physical ones, low-energy cutoff BL Lacs (LBLs) and
high-energy cutoff BL Lacs (HBLs), respectively.

We suspect that, in the $\alpha _{rx}-\log \nu _p$ plane, while probably all
the RBL-like objects are inside the upper left region and all the XBL-like
objects are confined in the lower right region, those in the upper right
region or lower left region may be: (1) (if it exists) the intermediate BL
Lacertae objects (IBLs) connecting RBL-like objects and XBL-like objects;
(2) the sources with their parabolic lines not passing through all the
values of $f_r$, $f_o$ and $f_x$ (as those shown in Figure 2); (3) the
sources with some of their fluxes being badly measured; (4) some extreme BL
Lacertae objects; (5) the sources not belonging to BL Lacertae objects.

\newpage
\begin{flushleft}
{\bf {\Large References}}\\
\end{flushleft}

\begin{verse}
Giommi, P., Ansari, S. G., \& Micol, A. 1995, A\&AS, 109, 267\\ Giommi, P.,
Barr, P., Garilli, B., Maccagni, D., \& Pollock, A. M. T. 1990, ApJ, 356, 432%
\\ Jannuzi, B. T., Smith, P. S., \& Elston, R. 1994, ApJ, 428, 130\\ Lamer,
G., Brunner, H., \& Staubert, R. 1996, A\&A, 311, 384\\ Ledden, J. E., \&
O'Dell, S. L. 1985, ApJ, 298, 630\\ Padovani, P., \& Giommi, P. 1995, ApJ,
444, 567\\ Padovani, P., \& Giommi, P. 1996, MNRAS, 279, 526\\ Perlman, E.
S., \& Stocke, J. T. 1993, ApJ, 406, 430\\ Sambruna, R. M., Maraschi, L., \&
Urry, C. M. 1996, ApJ, 463, 444\\ Stocke, J. T., Liebert, J., Schmidt, G.,
Gioia, I. M., Maccacaro, T., Schild, R. E., Maccagni, D., \& Arp, H. C.
1985, ApJ, 298, 619\\ Ulrich, M.-H., Maraschi, L., \& Urry, C. M. 1997,
ARA\&A, 35, 445\\ Urry, C. M., \& Padovani, P 1995, PASP, 107, 803\\
\end{verse}

\vspace{10mm} 
\begin{flushleft}
{\bf {\Large Figure caption:}}\\
\end{flushleft}

\begin{verse}
Figure 1: The $\alpha _{ro}-\alpha _{ox}$ diagram of BL Lacertae objects for
the sample of Giommi et al. (1995). Open circles represent the $36$ sources
with $2.4\times 10^{17}Hz\leq \nu _p\leq 4.8\times 10^{17}Hz$, open squares
represent the $7$ sources with $5.0\times 10^{12}Hz\leq \nu _p\leq 2.5\times
10^{13}Hz$, and filled diamonds represent the $60$ sources with $5.5\times
10^{14}Hz\leq \nu _p\leq 6.8\times 10^{14}Hz$. The solid line is the $\alpha
_{rx}=0.75$ line.

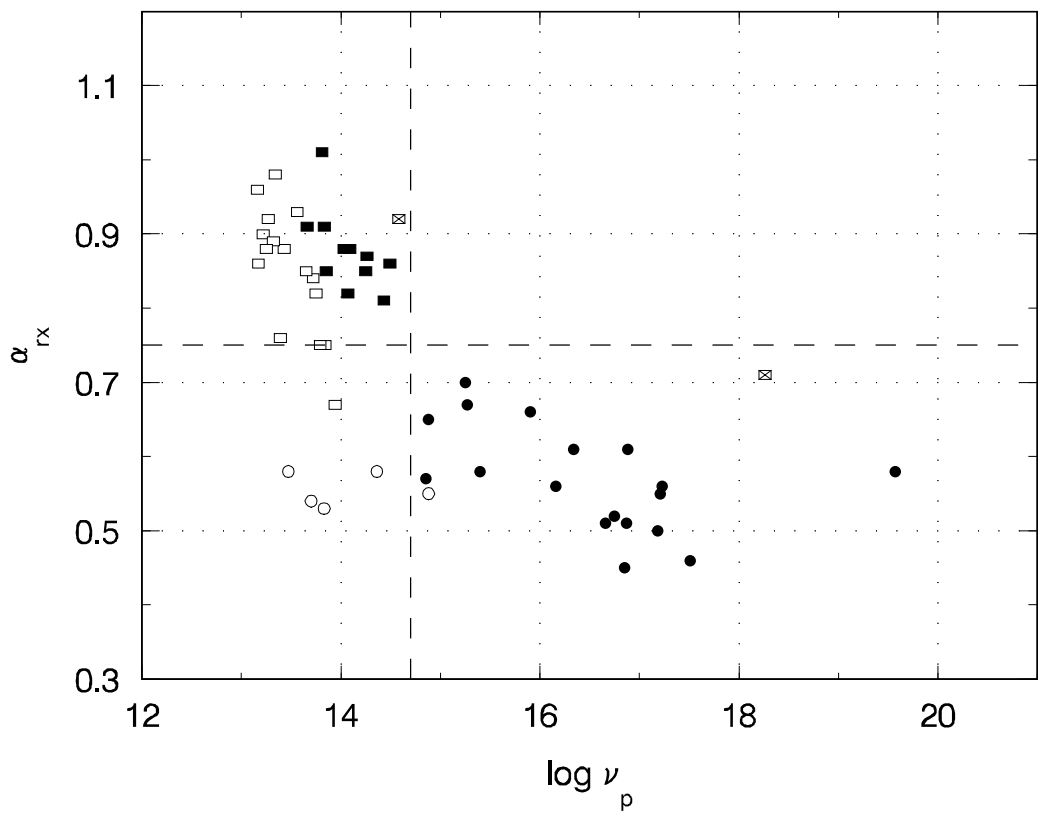
Figure 2: The plot of $\alpha _{rx}-\log \nu _p$ for the RBL and XBL samples
presented by Sambruna et al. (1996). Circles represent XBLs and squares
represent RBLs of the samples. Represented by empty symbols are the sources
with their parabolic lines being under the X-ray data points. The empty plus
cross symbols stand for the sources with their parabolic lines being above
the X-ray data points. The dashed horizontal line is the $\alpha _{rx}=0.75$
line and the dashed vertical line is the $\log \nu _p=14.7$ line.
\end{verse}

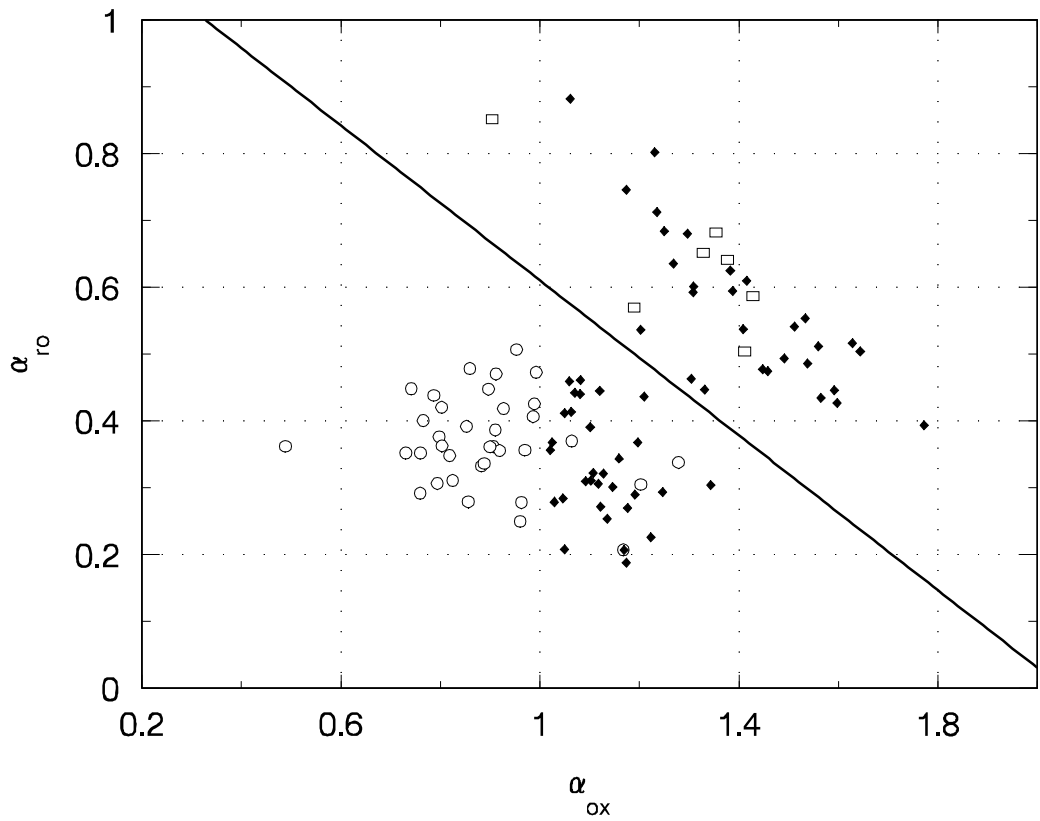
\end{document}